\documentclass[aps,physrev,twocolumn,groupedaddress]{revtex4-2}

\usepackage{siunitx}
\usepackage{graphicx}
\usepackage{amsmath}
\usepackage{subcaption}
\usepackage{placeins}
\usepackage{orcidlink}

\begin{document}


\title{\textbf{Impact of Rapid Acceleration on Beam Dynamics in the Rapid-Cycling Synchrotrons of a Muon Collider}}


\author{Lisa Soubirou\orcidlink{0009-0001-6491-988X}}

\author{Antoine Chance\orcidlink{0000-0002-6284-0086}}
\email[]{antoine.chance@cea.fr}

\affiliation{Universit\'e Paris-Saclay - CEA, France}


\date{\today}

\begin{abstract}
Circular muon colliders offer a promising route to multi-TeV centre-of-mass energies with high luminosity. The baseline design for the high-energy complex includes a chain of four Rapid-Cycling Synchrotrons (RCSs) accelerating the beam from \qty{63}{\GeV} to \qty{5}{\TeV}. The chain combines a conventional synchrotron with three hybrid synchrotrons, which use fixed-field superconducting dipoles and rapidly ramped normal-conducting dipoles. The short muon lifetime (\qty{2.2}{\micro\second} in the rest frame) imposes stringent constraints on the acceleration process: achieving the target luminosity requires a muon survival rate of up to \qty{70}{\percent} throughout the acceleration chain, with the full acceleration completed in approximately \qty{10}{\milli\second}. These demanding acceleration conditions raise several beam-dynamics challenges, investigated in this work. First, the impact of the RF cavity distribution on transverse emittance growth is studied. Second, the mismatch between the discretely increasing beam energy and the continuously ramped fields of the pulsed dipoles and quadrupoles is modelled analytically, and its impact is evaluated through tracking simulations. Finally, the sensitivity of the beam trajectory, optical functions, and emittance growth to magnetic-field ramp jitter is assessed.
\end{abstract}


\maketitle

\section{Introduction}

\begin{figure*}
\includegraphics[width=\textwidth]{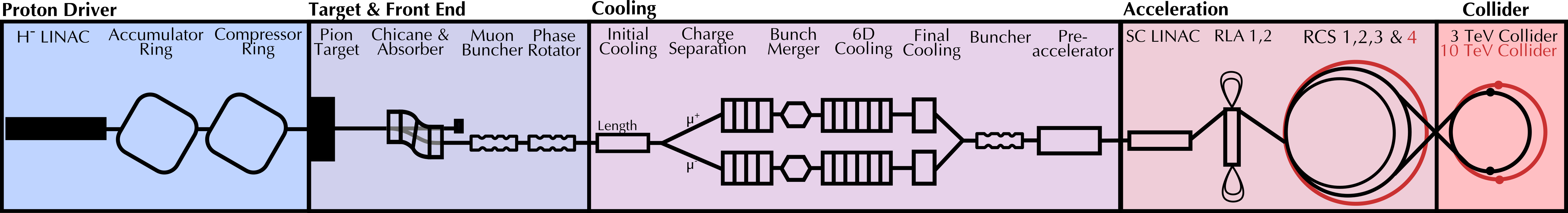}
\caption{Muon collider facility layout. The first stage uses three RCSs for a \qty{3}{\TeV} collider, with RCS 4 (in red) added in a second stage to reach \qty{10}{\TeV} \cite{consolidated_parameters}.\label{fig:muco_layout}} 
\end{figure*}

A muon collider operating at a centre-of-mass energy of \qty{10}{\TeV} and a luminosity of \qty{18e34}{\per \square \cm \per \s} would offer a rich physics programme \cite{IMCC_interimreport2025}. It would extend the discovery reach, enable precision studies of the Higgs boson, and probe new physics coupled to muons. The intense neutrino beams produced by muon decays would also provide opportunities for unprecedented studies \cite{ESPP2022, IMCC_interimreport2025}. A baseline design for a muon collider was developed by the US Muon Accelerator Program (MAP) \cite{MAP}. Building on this work, an updated design is currently under investigation within the International Muon Collider Collaboration (IMCC) and the MuCol programme \cite{IMCC_interimreport2025}. Its layout is shown in Fig.~\ref{fig:muco_layout}.  \par 

Following production, the muon beam undergoes cooling to reduce its transverse and longitudinal normalized emittances to \qty{25}{\micro\meter} and \qty{0.025}{\electronvolt\second}, respectively. The beam is then accelerated to \qty{63}{\GeV} using a linear accelerator followed by two recirculating linacs. A chain of four Rapid-Cycling Synchrotrons (RCSs) subsequently accelerates the beam before injection into the collider ring, where the two beams collide at a centre-of-mass energy of up to \qty{10}{\TeV}. The first RCS is a conventional synchrotron, while the remaining three employ a hybrid design combining fixed-field superconducting dipoles with rapidly ramped normal-conducting dipoles. The pulsed dipoles enable rapid acceleration, while the fixed-field superconducting dipoles provide a high average bending field, and help reduce the ring circumference. \par
The short muon lifetime, \qty{2.2}{\us} in the rest frame, is one of the main challenges for a muon collider, as production, cooling, acceleration, and collision must all take place within the relativistically dilated muon lifetime. Achieving the target luminosity requires a muon survival rate of up to \qty{70}{\percent} throughout the acceleration chain. The high-energy complex must therefore accelerate two counter-rotating single-bunch beams of $\mu^+$ and $\mu^-$, each containing approximately \num{2e12} muons, from \qty{63}{\GeV} to \qty{5}{\TeV} in approximately \qty{10}{\milli\second}, at a repetition rate of \qty{5}{\hertz}. The four RCSs operate with intermediate extraction energies of \qty{314}{\GeV}, \qty{750}{\GeV}, and \qty{1.5}{\TeV}. The main parameters governing this rapid acceleration are summarised in Table~\ref{tab:rcs_para_fast}. \par
\begin{table}
\caption{\label{tab:rcs_para_fast}Main parameters characterizing the fast acceleration of the four RCSs.}
\begin{ruledtabular}
\begin{tabular}{ccccc}
Parameter & RCS\,1 & RCS\,2 & RCS\,3 & RCS\,4 \\
\hline
C [\unit{\meter}] & 5990 & 5990 & 10700 & 35000 \\
$E_{\text{inj}}-E_{\text{ext}}$ [\unit{\giga \electronvolt}]
& 63--314 & 314--750 & 750--1500 & 1500--5000 \\
$t_{\text{acc}}$ [\unit{\milli \second}]
& 0.34 & 1.10 & 2.37 & 6.37 \\
$n_{\text{turns}}$
& 17 & 55 & 66 & 55 \\
$\Delta E_{\text{turn}}$ [\unit{\giga \electronvolt}]
& 14.8 & 7.9 & 11.4 & 63.6 \\
$V_{\text{tot}}$ [\unit{\giga \volt}]
& 27.6 & 17.5 & 15.7 & 72.7 \\
$G_\text{acc}$ [\unit{\mega \volt / \meter}]
& 2.44 & 1.33 & 1.06 & 1.83 \\
$\dot{B}$ [\unit{\tesla / \second}]
& 4232 & 3272 & 1519 & 565 \\
\end{tabular}
\end{ruledtabular}
\end{table}
The acceleration is particularly demanding in RCS1, where the beam energy increases from \qty{63}{\GeV} to \qty{314}{\GeV} in just \num{17} turns. This corresponds to an acceleration time of \qty{0.34}{\milli\second} and a peak magnetic-field ramp rate of \qty{4.2}{\kilo\tesla/\s}. Such rapid acceleration raises several beam-dynamics challenges, which are investigated in this work. First, the impact of RF cavity distribution on beam dynamics is studied. Second, the mismatch between the discretely increasing beam energy and the continuously varying fields of the pulsed dipoles and quadrupoles is described using an analytical model. Its impact on the beam trajectory, optical functions, and emittance growth is then evaluated. The investigation focuses initially on RCS1, the fastest accelerator in the chain, before extending to the remaining RCSs. Finally, the impact of a magnetic ramp jitter on the beam trajectory, optical functions, and emittance growth is investigated.

\section{RCS1 optics design\label{sec:rcs1_design}}
Two lattice designs are considered, with \num{8} and \num{32} arcs, respectively. Both feature \qty{90}{\degree} FODO cells and dispersion suppressors at the entrance and exit of the RF insertions. No chromaticity correction is implemented, and the lattices include only dipoles and quadrupoles. This choice is motivated by the compactness of the RCSs and the short acceleration cycle, lasting only a few tens of turns. Furthermore, transverse stability studies suggest that the beam can remain stable without chromaticity correction for orbit offsets below \qty{500}{\micro \meter} \cite{transverse_stab_imcc2026}. Let $\alpha_c$, $Q_{xys}$ and $\xi_{xy}$ be the momentum compaction, horizontal, vertical and longitudinal tunes, and horizontal and vertical chromaticities, respectively. The main parameters of the two lattice designs are summarised in Table~\ref{tab:rcs1_lattice_para}.

\begin{table}
\caption{\label{tab:rcs1_lattice_para}RCS1 lattice parameters at injection (inj.) and extraction (ext.) for 8 and 32 arcs.}
\begin{ruledtabular}
\begin{tabular}{ccccc}
 & \multicolumn{2}{c}{8-arc lattice} & \multicolumn{2}{c}{32-arc lattice} \\
Parameter & 1 arc & Full RCS & 1 arc & Full RCS \\
\hline
$\alpha_c$ & \multicolumn{2}{c}{\num{7.89e-4}} & \multicolumn{2}{c}{\num{6.78e-4}} \\
$Q_s$(at inj.) & 0.134 & 1.068 & 0.030 & 0.963 \\
$Q_s$(at ext.) & 0.059 & 0.468 & 0.013 & 0.431 \\
$Q_x$      & 5.129 & 41.035 & 1.314 & 42.053 \\
$Q_y$      & 5.080 & 40.637 & 1.116 & 35.702 \\
$\xi_x$    & -7.104 & -56.835 & -2.191 & -70.101 \\
$\xi_y$    & -6.418 & -51.341 & -1.494 & -47.807 \\
\end{tabular}
\end{ruledtabular}
\end{table}

\begin{figure}
\includegraphics[width=0.9\columnwidth]{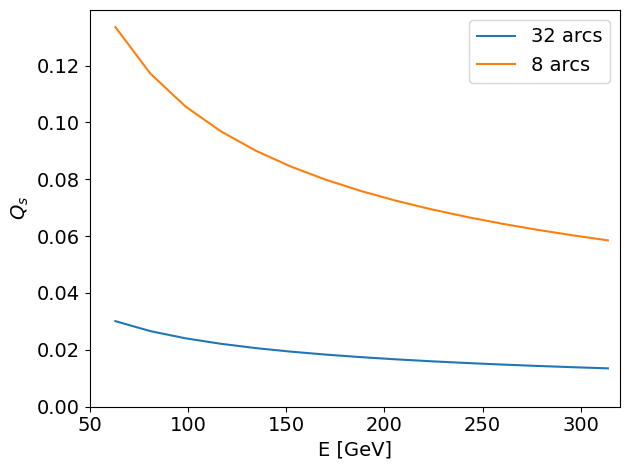}
\caption{Variation of the synchrotron tune $Q_s$ per super-period as function of the beam energy, for RCS1 8-arc and 32-arc lattices. \label{fig:rcs1_qs_var}} 
\end{figure}

As minimising the installed RF voltage requires the most linear acceleration ramp possible, a constant energy gain per turn is assumed \cite{Batsch:2886626}. However, the very short acceleration times, comparable to the cavity filling time \cite{thiele2025beamcavity}, make it difficult to adjust the RF voltage dynamically throughout the ramp. The RF voltage is therefore kept constant during acceleration, causing the synchrotron tune to evolve significantly over the acceleration cycle. The synchrotron tune evolution per super-period is shown in Fig.~\ref{fig:rcs1_qs_var} for both lattice configurations. The tune per super-period varies from approximately \num{0.13} to \num{0.06} for the 8-arc lattice and from \num{0.03} to \num{0.01} for the 32-arc lattice. Although the global synchrotron tune remains similar in both configurations, the longitudinal phase advance between consecutive RF sections is larger in the 8-arc lattice. Despite the relatively distributed RF system compared with a conventional synchrotron, the 8-arc lattice operates close to the stability limit of $1/\pi$ \cite{synchro_beta_suzuki}. This limit defines the maximum phase advance between RF stations for stable synchrotron motion and phase stability when the RF system is concentrated in a small number of stations. These differences motivate a detailed investigation of the impact of RF cavity distribution on beam dynamics.

\section{Transverse emittance growth in RCS1: impact of RF distribution}

\begin{figure}
\includegraphics[width=0.9\columnwidth]{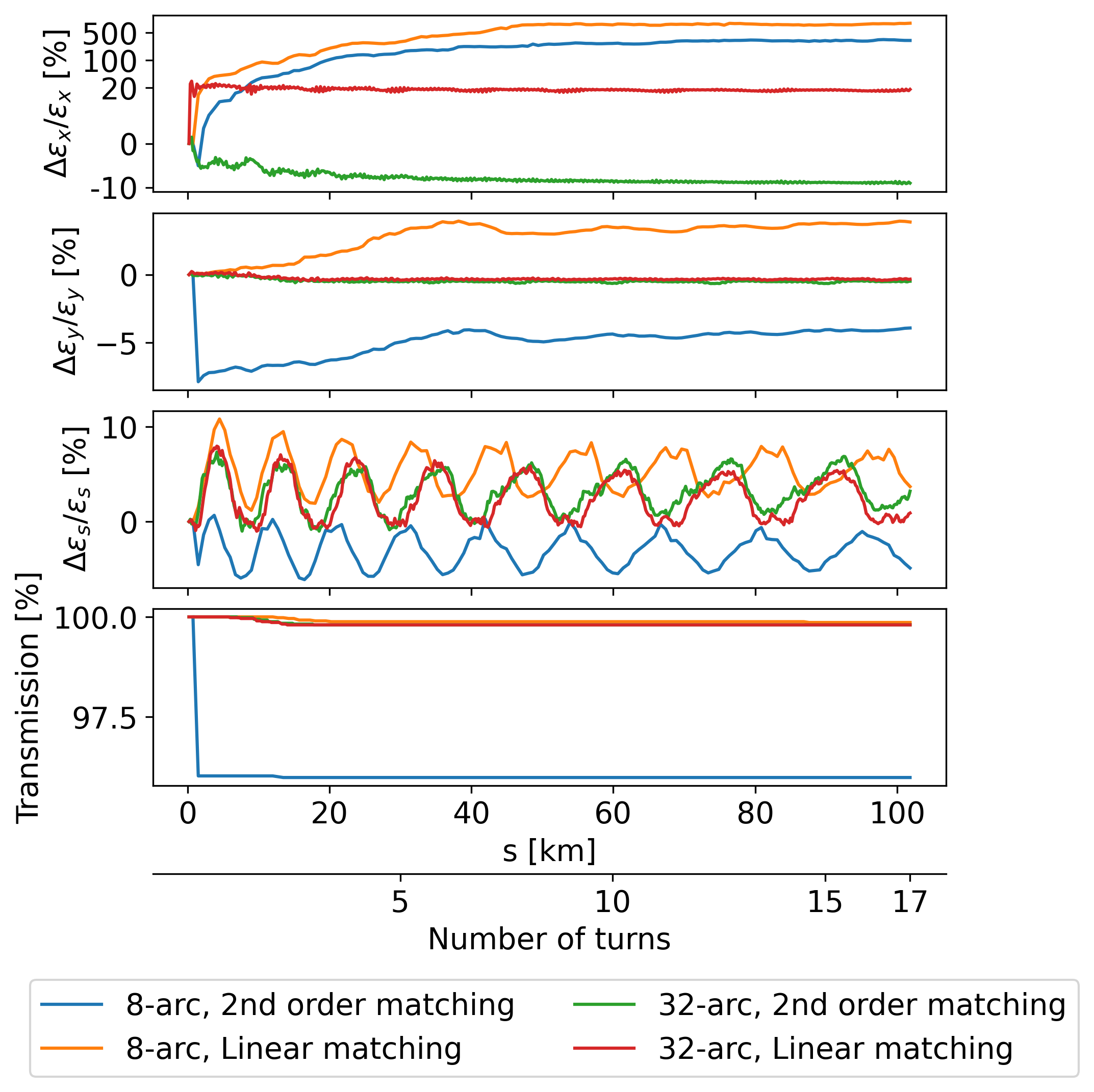}
\caption{Tracking srudies for the RCS1 8-arc and 32-arc lattices, with initial particle distribution linearly matched or matched at second order.\label{fig:tracking_rcs1_matching}} 
\end{figure}

Tracking studies are performed for both the 8-arc and 32-arc lattices using matched beam distributions in the transverse and longitudinal planes. The results are presented in Fig.~\ref{fig:tracking_rcs1_matching}. In the 8-arc lattice, the horizontal emittance $\varepsilon_x$ increases dramatically, with a final emittance growth exceeding \qty{500}{\percent}. In contrast, the 32-arc lattice exhibits a much smaller horizontal emittance growth of \qty{13}{\percent}. \par

When the injected beam is additionally matched to second order in the transverse plane, the horizontal emittance growth in the 32-arc lattice is completely suppressed. The slight decrease in emittance observed towards the end of the tracking results from particle losses. As the longitudinal emittance approaches the RF bucket acceptance, a small fraction of the injected particles falls outside the RF acceptance, as shown in Fig.~\ref{fig:rcs1_es_part0}. Applying the same second-order matching procedure to the 8-arc lattice slightly improves the beam behaviour, but remains largely insufficient to reduce the emittance growth to acceptable levels. A few percent of the particles are lost during tracking, indicating that the second-order matching is less effective for this lattice. Several tune settings were then explored to assess their impact on emittance growth. In particular, the arc lattice was carefully tuned to shift the fractional horizontal tune away from the longitudinal tune at injection. However, this did not lead to any improvement. This may be because the synchrotron tune varies rapidly during acceleration, which could prevent resonance excitation. Other lattice configurations were also considered by varying the number of FODO cells in the arcs, thereby exploring different fractional tune values. Nevertheless, none of the configurations tested reduced the horizontal emittance growth to an acceptable level.\par

These results highlight the importance of carefully designing the injection optics to minimise the Montague functions $W_{x,y}$ and the second-order dispersion $D_{2,x}$. They also show that the distribution of the RF cavities around the ring has significantly impacts the transverse beam dynamics, suggesting the presence of synchro-betatron coupling. A more detailed investigation was therefore carried out using the 8-arc lattice of RCS1.

\subsection{Influence of the longitudinal emittance}

\begin{figure}
\includegraphics[width=0.9\columnwidth]{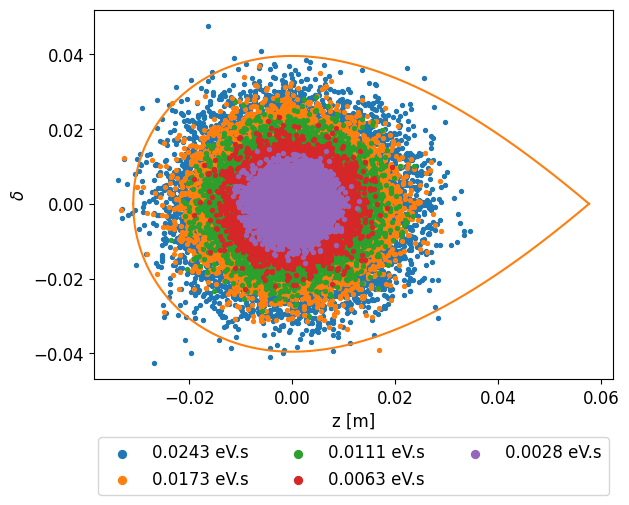}
\caption{Initial longitudinal particle distribution for varying initial longitudinal emittance and RF bucket separatrix at injection into RCS1. \label{fig:rcs1_es_part0}}
\end{figure}

\begin{figure}[!htbp]
        \centering
        \includegraphics[width=0.9\columnwidth]{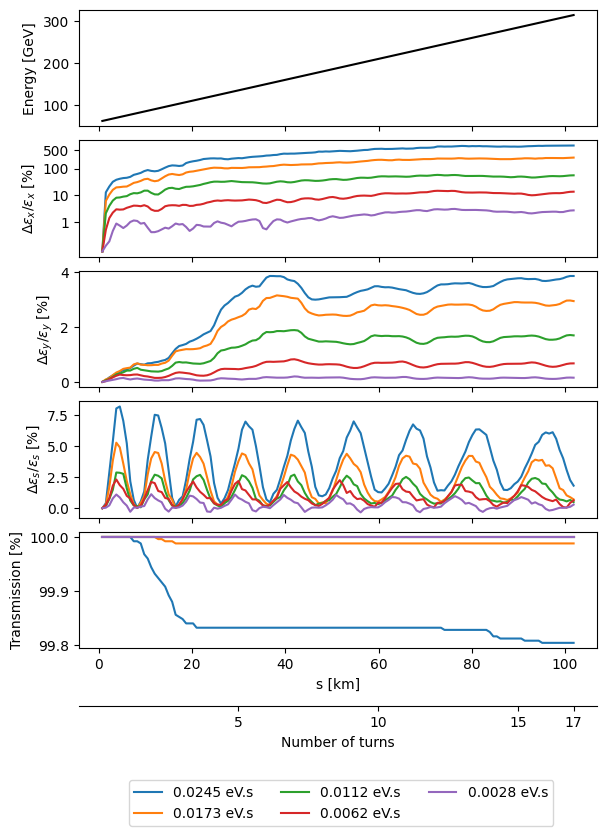}
        \caption{Tracking studies for the 8-arc lattice of RCS1 with decreasing initial longitudinal emittance. \label{fig:rcs1_tracking_es}}
    \end{figure}

\begin{figure}
        \centering
       \includegraphics[width=0.9\columnwidth]{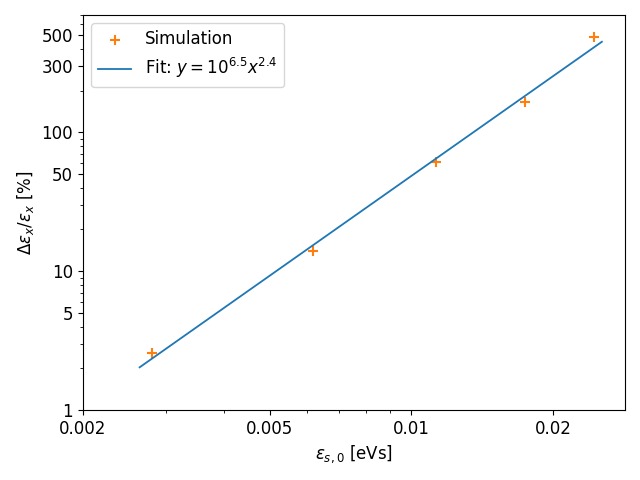}
        \caption{Final horizontal emittance growth  for the 8-arc lattice of RCS1 as a function of the
initial longitudinal emittance.}
        \label{subfig:rcs1_tracking_es_emit}
\end{figure}

First, a scan of the longitudinal emittance is performed to investigate its impact on the transverse beam dynamics and assess the presence of synchro-betatron coupling. To reduce the longitudinal emittance, different initial bunch lengths ranging from \SIrange{10}{30}{\pico\second} are considered. The corresponding beam distributions, with longitudinal emittances ranging from the nominal value of \qty{0.025}{\electronvolt\second} down to \qty{0.0028}{\electronvolt\second}, are then tracked through the lattice. The initial longitudinal particle distributions, together with the RF bucket separatrix at injection in RCS1, are shown in Fig.~\ref{fig:rcs1_es_part0}. The tracking results presented in Fig.~\ref{fig:rcs1_tracking_es} show a strong reduction in horizontal emittance growth as the longitudinal emittance decreases. The growth decreases from more than \qty{500}{\percent} for the nominal longitudinal emittance to only a few percent for a longitudinal emittance of \qty{0.0028}{\electronvolt\second}. This demonstrates a strong correlation between the longitudinal and transverse emittance evolution. Over the range of longitudinal emittances investigated, this dependence follows a power-law behaviour, as shown in Fig.~\ref{subfig:rcs1_tracking_es_emit}.

An interesting behaviour is also observed in the longitudinal plane. The longitudinal emittance exhibits oscillations whose amplitude decreases as the initial longitudinal emittance is reduced. These oscillations are attributed to a longitudinal mismatch at injection. Although the RF voltage provides a large bucket acceptance, the large longitudinal emittance causes a significant fraction of the beam distribution to extend close to or beyond the separatrix. Particles with large longitudinal amplitudes therefore leave the linear regime and no longer follow elliptical trajectories in longitudinal phase space. Their nonlinear synchrotron motion leads to a temporary increase in the longitudinal emittance. After approximately one synchrotron period, the emittance returns to its initial value. Since the synchrotron tune evolves throughout the acceleration cycle, the oscillation frequency also changes with time.

\subsection{Influence of the injection energy}

\begin{figure}
\includegraphics[width=0.9\columnwidth]{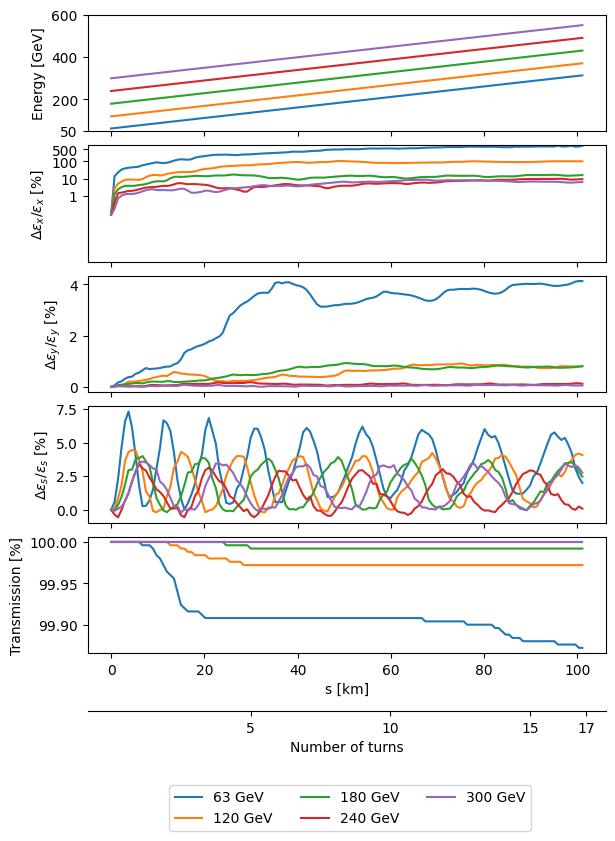}
\caption{Tracking studies for the 8-arc lattice of RCS1 with increasing injection energies. \label{fig:rcs1_tracking_energy}} 
\end{figure}

The second step investigates the effect of a fixed acceleration rate for different injection energies. The same lattice is tracked in all cases, using an identical energy gain per turn and number of turns. Consequently, varying the injection energy results in different extraction energies. The objective is to assess the sensitivity of the emittance growth to the acceleration rate as a function of the injection energy.
The injection energy is scanned from \qty{63}{\giga\electronvolt} to \qty{300}{\giga\electronvolt}, with the corresponding results presented in Fig.~\ref{fig:rcs1_tracking_energy}. The final horizontal emittance growth decreases progressively with increasing injection energy. For the current lattice configuration, the emittance growth is reduced below \qty{20}{\percent} for injection energies above approximately \qty{180}{\giga\electronvolt}. Further reduction would require, as discussed previously, optimisation of the second-order optical functions. Longitudinal emittance oscillations are also observed during acceleration for all injection energies, but with different frequencies. This behaviour is consistent with the decrease in synchrotron tune as the beam energy increases, which leads to slower synchrotron oscillations at higher energies.
These results indicate that the sensitivity of RCS1 to fast acceleration is partly driven by its relatively low injection energy, which results in larger relative energy variations during acceleration. This effect is not observed in the downstream RCSs, which operate at significantly higher injection energies.

\subsection{Conclusion}

In conclusion, a sparse distribution of the RF sections in RCS1 results in significant horizontal emittance growth, revealing a strong synchro-betatron coupling. This effect is greatly reduced when the initial longitudinal emittance is decreased, showing that particles with large longitudinal amplitudes drive the emittance growth. For the nominal longitudinal emittance required by the accelerator, the RF sections must be more evenly distributed around the ring, requiring 32 RF insertions to preserve the beam quality. However, it is worth noting that distributing and synchronising RF power across a large number of RF stations represents a significant technological challenge. Even with this configuration, the relatively low injection energy of RCS1 makes the beam particularly sensitive: a careful optimisation of the second-order optical functions at injection is therefore required to fully suppress the remaining emittance growth.

\section{Effects of energy–magnetic field mismatch}

In a conventional accelerator, where the acceleration is adiabatic, the variation of the magnetic field along the arcs during the passage of the beam can be neglected. This approximation no longer holds in the RCS. It is therefore critical to distribute the RF system: increasing the number of RF sections reduces the energy gain per section, and consequently the energy mismatch between the beam and the magnetic field. To minimise this mismatch, the magnetic field ramp was phase-shifted with respect to the beam energy such that the field is matched to the beam energy at the centre of each arc. In this configuration, it represents a dipolar error up to \qty{1.5}{\percent} in RCS1, with \num{8} RF sections. The following discusses the development of an analytical model and investigates the impact of mismatch on beam trajectory, optics and emittance growth.

\subsection{Analytical model}

An analytical model is developed to describe the varying magnetic field within a pulsed element. We consider an arc parametrized by the longitudinal coordinate $s$, with $s=0$ defined at the beginning of the arc. Let a pulsed dipole of length $L$ be located such that its entrance face is at position $s_0$. Due to the magnetic field ramp, the strength of an infinitesimal slice of dipole depends on its longitudinal position $s$ along the arc and can be expressed as
\begin{equation}
    k_0(s) = k_0^* + k'_0 \ ( s - s^*)
\end{equation}    
with
\begin{equation}
    k'_0 \equiv \frac{1}{B \rho} \frac{\dot{B}}{c} 
\end{equation}
Here, $k_0^*$ denotes the dipole strength at the reference energy, $s^*$ the longitudinal position at which the magnetic field reaches its reference value, $\dot{B}$ the dipole ramp rate, ($B \rho$) the beam magnetic rigidity, and $c$ the speed of light.

Similarly, for a pulsed quadrupole, the gradient varies along the arc according to
\begin{equation}
    k_1(s) = k_1^* + k'_1 \ ( s - s^*)
\end{equation}
with
\begin{equation}
    k'_1 \equiv \frac{1}{B \rho} \frac{\dot{G}}{c}
\end{equation}
where $k_1^*$ is the quadrupole strength at the reference energy and $\dot{G}$ the quadrupole ramp rate. \par
In the following, $M_{a \rightarrow b}$ and $T_{a \rightarrow b}$ denote the reference first order and second-order transfer matrices from $s=a$ to $s=b$. We define the element reference transport matrix as $M=M_{s_0 \rightarrow s_0+L}$. The quantities $\delta X(s)$ and $\delta M(s)$ denote the infinitesimal perturbations of the offset and first order matrix in $s$. Finally, $\Delta X$ and $\Delta M$ are the integrated offset and matrix perturbation of the element.
Within this framework, the transfer matrix of a pulsed element can be expressed as the sum of the reference matrix evaluated at the nominal energy and a perturbative correction:
\begin{equation}
    M_{\text{pulsed}} = M + \Delta M
\end{equation}

The integrated orbit offset $\Delta X$ and the matrix perturbation $\Delta M$ for a pulsed dipole can then be evaluated consistently up to second order as
\begin{align}\Delta X(s_0) & = \int_{s_0}^{s_0+L} M_{s \rightarrow s_0+L} \, \delta X(s) \, ds \\
\Delta M(s_0) &= \int_{s_0}^{s_0+L} M_{s \rightarrow s_0+L} \, \delta M(s) \, M_{s_0 \rightarrow s} \, ds \notag \\
    &\quad + 2 \int_{s_0}^{s_0+L} T_{s \rightarrow s_0+L}\bigl(M_{s_0 \rightarrow s}, \delta X(s)\bigr) \, ds
\end{align}
The local perturbations depend on the type of pulsed element. For a pulsed dipole, the only non-zero terms of $\delta X(s)$ and $\delta M(s)$ are:
\begin{align}
    \delta X_{\text{dip}\,2}(s) &= - k'_0 (s-s^*) \\
    \delta M_{\text{dip}\,2,1}(s) &= -k_0^* \ k'_0 (s-s^*)
\end{align}
In contrast, a pulsed quadrupole does not generate an orbit offset, and only two terms of the matrix perturbation are non-zero:
\begin{align}   
    \delta X_{\text{qd}}(s) &= 0 \\
    \delta M_{\text{qd}\, 2,1}(s) &= - k'_1 (s - s^*) \\
    \delta M_{\text{qd}\, 4,3}(s) &= k'_1 (s - s^*)
\end{align}

The analytical expressions for the trajectory offset and matrix perturbation induced by pulsed dipoles and quadrupoles are provided in Appendix~\ref{sec:appendix}. For the pulsed dipole, the resulting orbit offset is of particular interest. In the limit $k_0^* L \ll 1$, it reduces to
\begin{align}
\label{eq:dipole_offset_appr}
    \Delta X_{dip,1} &= - \dfrac{1}{6} \frac{1}{B \rho} \frac{\dot{B}}{c}  \, L^{2} \, (L + 3 s_0 - 3 s^*) \\
    \Delta X_{dip,2} &= -\dfrac{1}{2}  \frac{1}{B \rho} \frac{\dot{B}}{c} \, L \, (L + 2 s_0 - 2 s^*)\\
    \Delta X_{dip,5} &=\dfrac{1}{24} \frac{1}{B \rho} \frac{\dot{B}}{c}\dfrac{k_0^*}{\beta_0} L^3 \, (L + 4 s_0 - 4 s^*) \label{eq:dipole_offset_appr_zeta}
\end{align}
where $\beta_0$ is the reference relativistic velocity. It provides a simple expression for the offset induced by the field ramp in a dipole.

The analytical model characterises the perturbation introduced by pulsed elements through the integrated offset $\Delta X$ and the corresponding transfer matrix correction $\Delta M$. Since the transfer matrix of the full line is obtained through matrix multiplication, the perturbation is represented by an equivalent thin-lens matrix $M_\text{eq}$, defined as
\begin{equation}
    M_\text{eq} = I + \Delta M \cdot M^{-1}
\end{equation}
This equivalent matrix can be implemented as a thin-lens element placed immediately after each pulsed element. For comparison, a numerical model is implemented in \textsc{Xsuite} \cite{xsuite} by slicing each pulsed element and assigning a progressively varying magnetic field to each slice. This sliced model provides a numerical benchmark for the analytical model.

\subsection{Model validation}
\begin{figure}
    \begin{subfigure}{0.45\textwidth}
        \centering
        \includegraphics[width=\linewidth]{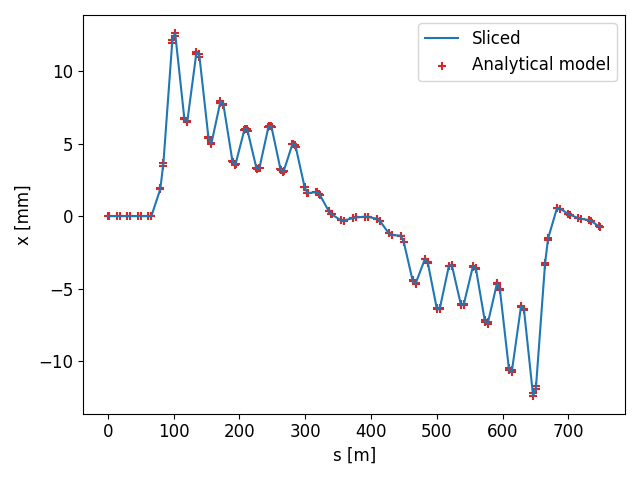}
        \caption{Horizontal coordinate $x$.}
        \label{subfig:rcs1_8arc_x}
    \end{subfigure}
    \begin{subfigure}{0.45\textwidth}
        \centering
        \includegraphics[width=\linewidth]{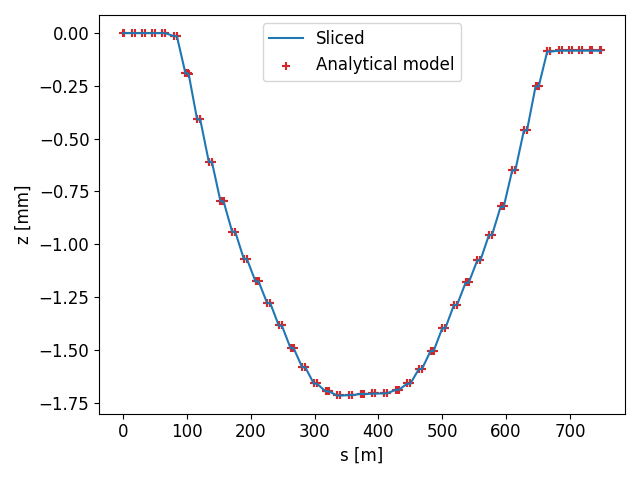}
        \caption{Longitudinal coordinate $z$.}
        \label{subfig:rcs1_8arc_z}
    \end{subfigure}
\caption{Trajectory perturbation in one super-period of the 8-arc lattice of RCS1. \label{fig:rcs1_8arc_traj}} 
\end{figure}

The benchmark is done on the 8-arc lattice. The trajectory offset caused by the energy-magnetic field mismatch is presented in Fig.~\ref{fig:rcs1_8arc_traj}. An excellent agreement is observed between the analytical model and the slicing, with discrepancies remaining at the micrometer level for the horizontal coordinate $x$. A similar study of the longitudinal coordinate $z$ shows comparable agreement. In the first half of the arc, the dipole magnetic field is weaker than the reference magnetic field, causing the beam to move outwards the ring. The resulting trajectory is longer than the reference trajectory, leading to a negative deviation of the longitudinal coordinate. In the second half of the arc, the dipole field becomes stronger than the reference field, steering the beam back inwards the ring and consequently reducing the path-length difference. Since the magnetic field matches the reference value at the centre of the arc, the trajectory deviations accumulated in the first half are compensated in the second half, and both the transverse and longitudinal coordinate deviations return close to zero by the end of the arc.

\begin{figure}
\includegraphics[width=0.9\columnwidth]{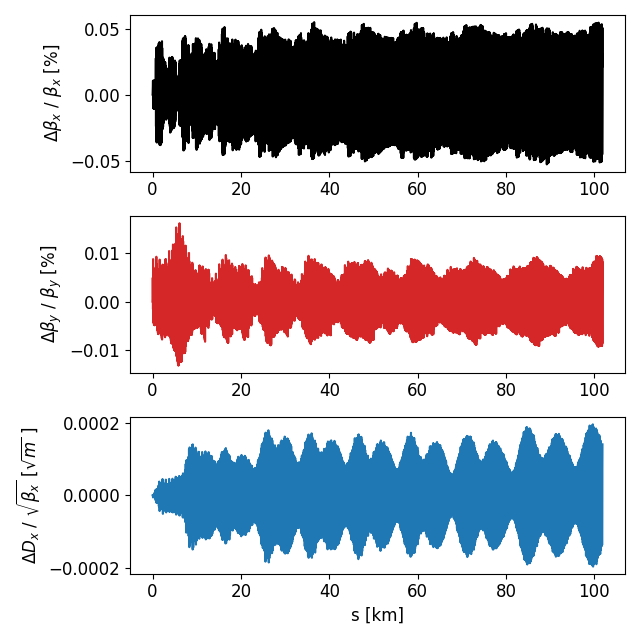}
\caption{Start-to-end optics perturbation difference between the analytical model and the slicing with respect to the reference optics for RCS1 8-arc lattice. \label{fig:rcs1_8arc_optics_difference}}
\end{figure}

We now turn to the optics perturbation. We take interest in the beta functions and the dispersion function. A comparison between the analytical model and the slicing was done. The agreement was very good, with a discrepancy below \qty{0.05}{\percent} for $\beta_x$, below \qty{0.01}{\percent} for $\beta_y$ and below \qty{0.0002}{\sqrt{\meter}} for the normalized dispersion, which all represent an error well below \qty{1}{\percent}, as shown in Fig.~\ref{fig:rcs1_8arc_optics_difference}.

\subsection{Start-to-end results}
\subsubsection{Trajectory}

\begin{figure}
    \begin{subfigure}{0.45\textwidth}
        \centering
        \includegraphics[width=\linewidth]{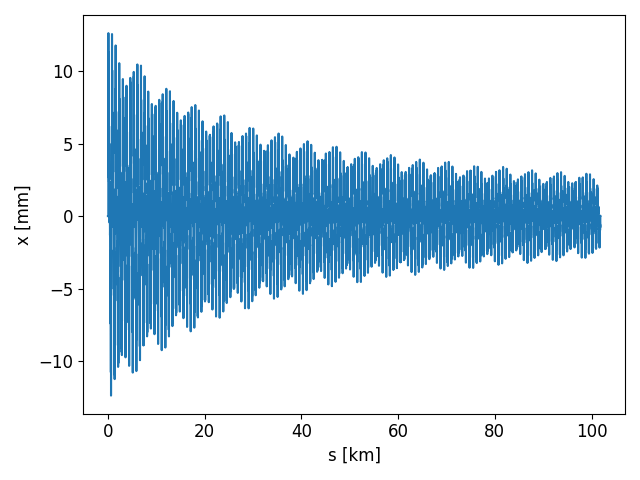}
        \caption{Horizontal coordinate x.}
        \label{subfig:rcs1_8arc_x_full}
    \end{subfigure}
    \begin{subfigure}{0.45\textwidth}
        \centering
        \includegraphics[width=\linewidth]{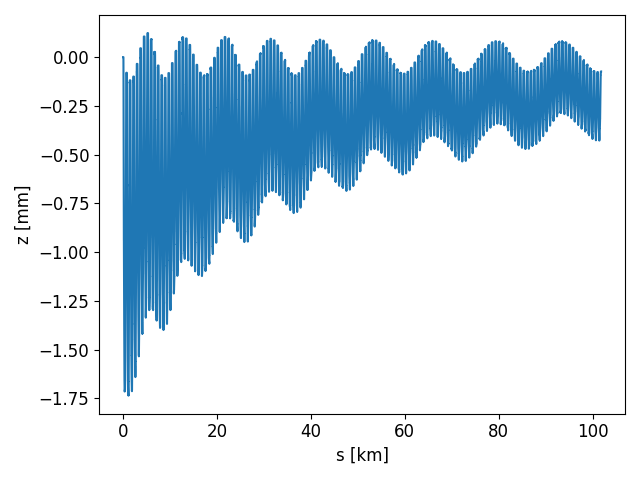}
        \caption{Longitudinal coordinate z.}
        \label{subfig:rcs1_8arc_z_full}
    \end{subfigure}
\caption{Trajectory perturbation from injection to extraction in the 8-arc lattice of RCS1. \label{fig:rcs1_8arc_traj_full}} 
\end{figure}

We now extend the study to the full machine, tracking the beam from injection to extraction by building the complete lattice. Figure~\ref{fig:rcs1_8arc_traj_full} shows the evolution of the horizontal coordinate $x$ and the longitudinal coordinate $z$ from start-to-end.

As shown in Fig.~\ref{subfig:rcs1_8arc_x_full}, the beam undergoes oscillations in each successive arc of the RCS, with a progressively decreasing amplitude and a maximum trajectory offset of approximately \qty{13}{\milli \meter}. This behaviour results from adiabatic damping as the beam energy increases, in agreement with the $1/(B\rho)$ scaling of the orbit offset induced by a pulsed dipole predicted in Eq.~\eqref{eq:dipole_offset_appr}.  As shown in Fig.~\ref{subfig:rcs1_8arc_z_full} and predicted by Eq.~\eqref{eq:dipole_offset_appr_zeta}, the variation of the longitudinal coordinate $z$ also decreases throughout the acceleration cycle. However, as Fig.~\ref{subfig:rcs1_8arc_z} shows, the longitudinal coordinate does not return to zero at the end of each arc, resulting in synchrotron oscillations. While a tuner compensates for the path-length variation due to the beam not being fully ultra-relativistic yet, no additional compensation is applied to account for the path-length mismatch induced by the dipole ramp. Nevertheless, the resulting relative energy deviation at extraction remains below \num{1e-4}.

\subsubsection{Optics perturbation}

\begin{figure}
\includegraphics[width=0.9\columnwidth]{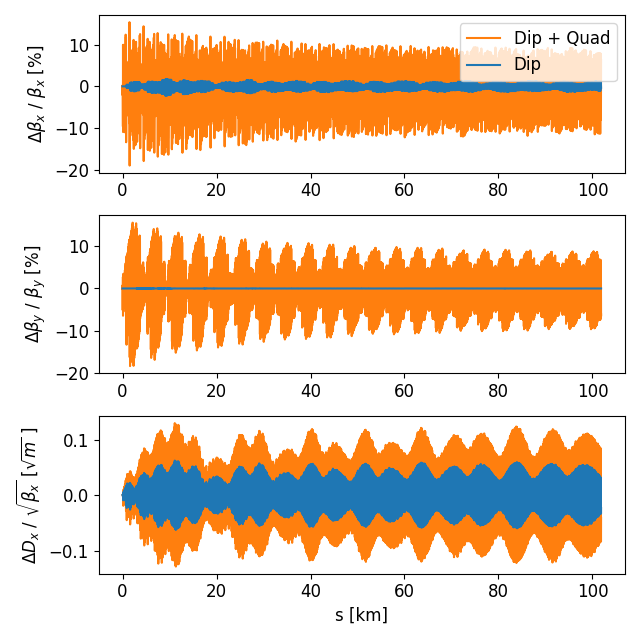}
\caption{Start-to-end optics perturbation with respect to the reference optics for RCS1 8-arc lattice. In blue, the energy-magnetic field is applied only to the dipoles; in orange, it is applied to both dipoles and quadrupoles. \label{fig:rcs1_8arc_optics}}
\end{figure}

To understand the contribution of each type of pulsed element, we compare two scenarios. In the first, the energy-magnetic field mismatch is applied only to the dipoles, while the quadrupoles remain perfectly matched to the beam energy. In the second, it is applied simultaneously to both the dipoles and the quadrupoles. The corresponding results are shown in Fig.~\ref{fig:rcs1_8arc_optics}. 

The dispersion perturbation is primarily driven by the dipoles and further amplified by the quadrupole mismatch. When the mismatch is applied to both dipoles and quadrupoles, the perturbation of the normalized dispersion reaches up to \qty{0.1}{\sqrt{\meter}}. In contrast, the $\beta$-beating is dominated by the quadrupoles, as the weak focusing provided by the dipoles is almost negligible in such a large machine. A $\beta$-beating of up to \qty{20}{\percent} is observed, which is particularly severe given that other sources of optics errors, such as field errors and alignment defects, have not yet been included in the model. The improvement achieved with the 32-arc lattice is quantified in Table~\ref{tab:EB_rcs_chain}, with a maximum trajectory offset of \qty{4.1}{\milli \meter} and a maximum horizontal $\beta$-beating of \qty{3.3}{\percent}.

\subsubsection{Tracking studies and emittance growth}

\begin{figure}
        \centering
        \includegraphics[width=\linewidth]{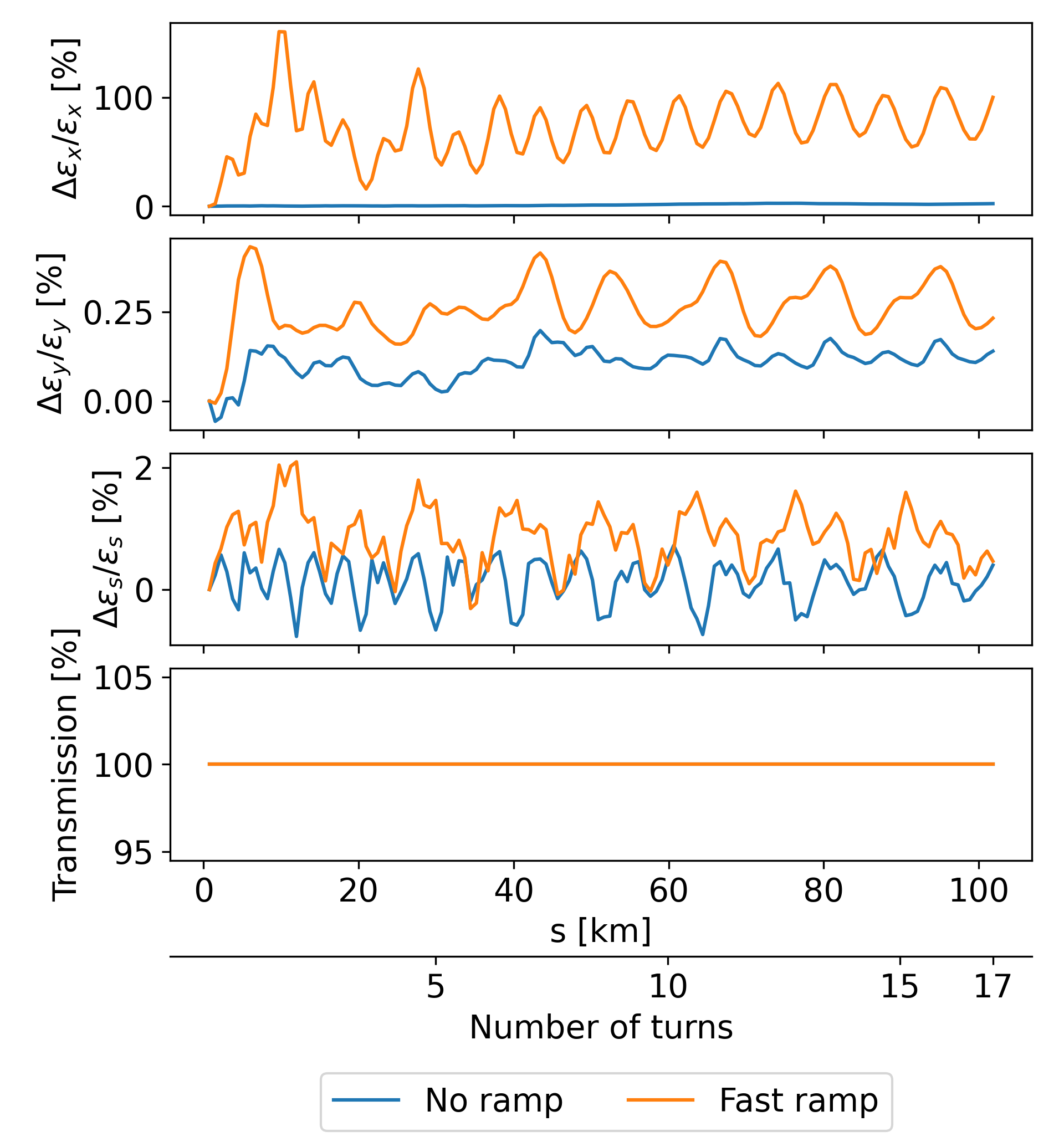}
\caption{Tracking studies for the 8-arc RCS1 lattice, with (Fast ramp) and without (No ramp) energy–magnetic-field mismatch. The longitudinal emittance is set to $\varepsilon_s=\qty{0.0028}{\electronvolt\second}$. \label{fig:track_EB_8}} 
\end{figure}

\begin{figure}
        \centering
        \includegraphics[width=\linewidth]{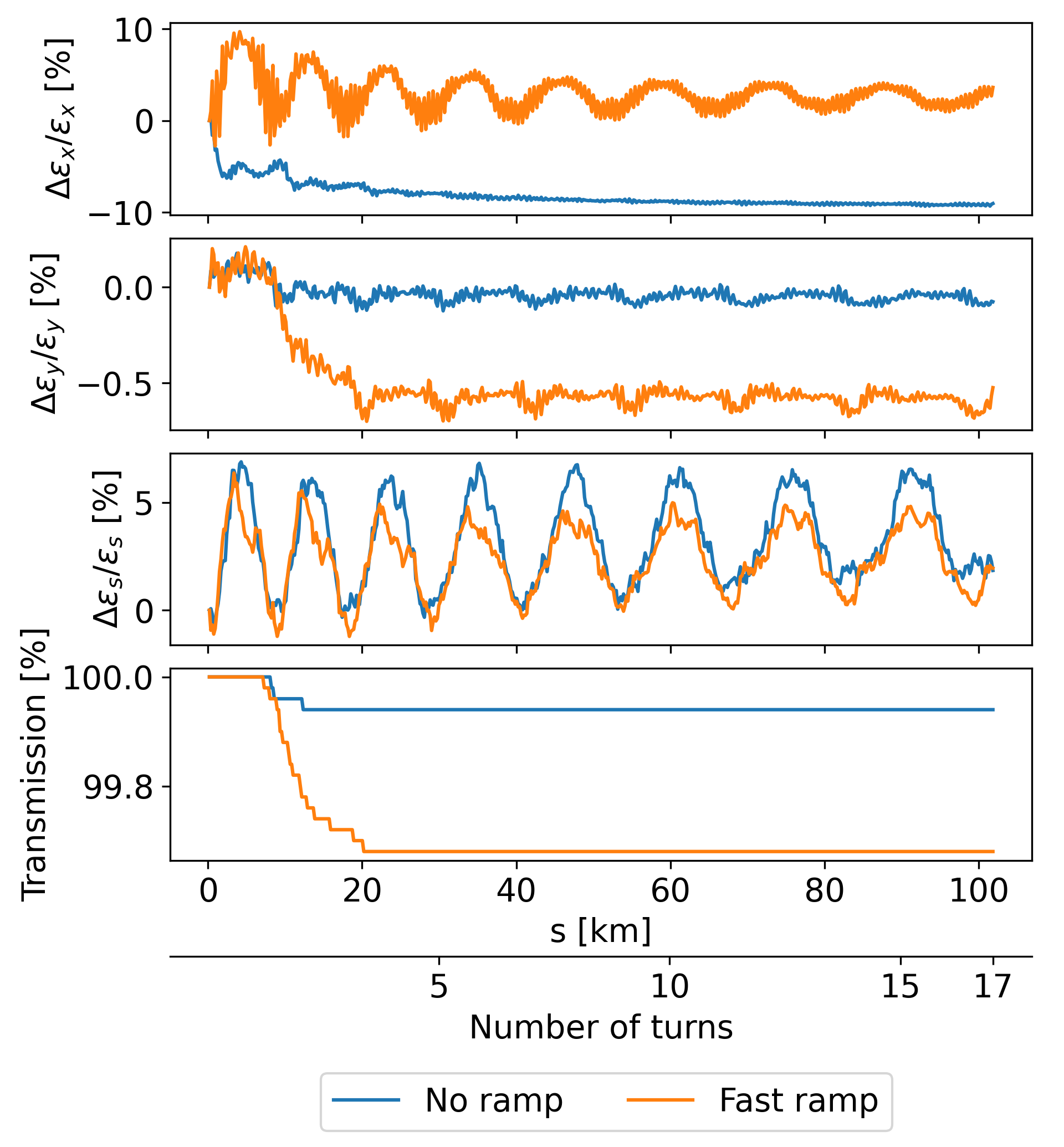}
\caption{Tracking studies for the 32-arc RCS1 lattice, with (Fast ramp) and without (No ramp) energy–magnetic-field mismatch. The nominal longitudinal emittance is used: $\varepsilon_s=\qty{0.025}{\electronvolt\second}$. \label{fig:track_EB_32}} 
\end{figure}

Tracking studies are performed on RCS1 using both the 8-arc and 32-arc lattices. Given the horizontal emittance growth observed for the 8-arc lattice with the nominal longitudinal emittance of $\varepsilon_s=\qty{0.025}{\electronvolt\second}$, the impact of the fast ramp is investigated using a reduced longitudinal emittance of $\varepsilon_s=\qty{0.0028}{\electronvolt\second}$. Despite this substantial reduction, the horizontal emittance growth in the 8-arc lattice still exceeds \qty{100}{\percent} when the energy–magnetic-field mismatch is included, as shown in Fig.~\ref{fig:track_EB_8}. In contrast, the 32-arc lattice, tracked with the nominal longitudinal emittance, exhibits an emittance growth of approximately \qty{7}{\percent}, as presented in Fig.~\ref{fig:track_EB_32}. The horizontal emittance oscillates at the same frequency as the longitudinal emittance, corresponding to the synchrotron tune, further highlighting the presence of synchro-betatron coupling. \par
These results further strengthen the need for a distributed RF system. Indeed, \num{8} RF sections are clearly insufficient, even with a longitudinal emittance reduced by almost one order of magnitude. The energy-magnetic field mismatch induced by the fast magnet ramp therefore places strong constraints on the RF section distribution to limit horizontal emittance growth.

\subsubsection{Impact of the ramp rate}

\begin{figure}
\includegraphics[width=0.9\columnwidth]{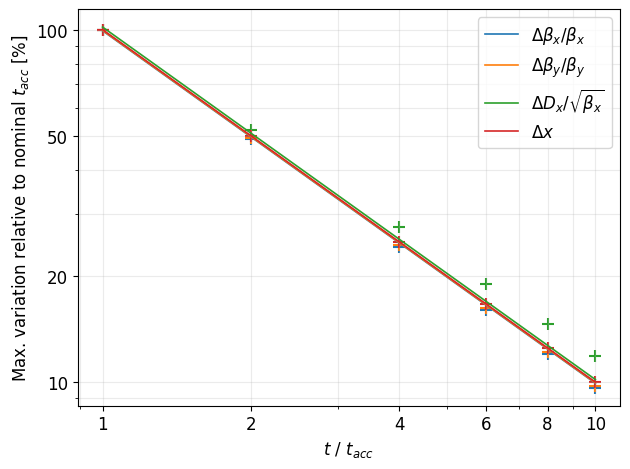}
\caption{RCS1 (8-arc lattice). Effect of the ramp speed on the relative variation of the optics functions and maximum orbit offset. The points correspond to the simulation results, while the solid lines show a $1/x$ fit. \label{fig:scan_tacc}}
\end{figure}

A key parameter governing this perturbation is the ramp rate. We therefore investigate how the perturbation scales with this parameter. Reducing the ramp rate is equivalent to accelerating the beam over the same energy range but with a longer acceleration time. We define a scaling factor as the ratio of the extended acceleration time to the nominal acceleration time.
The results are shown in Fig.~\ref{fig:scan_tacc}. As expected, both the relative variation of the optics functions and the maximum orbit offset decrease as the acceleration time increases. More specifically, both quantities exhibit an inverse proportionality with the acceleration time, confirming that the perturbation is proportional to the ramp rate, as predicted by the analytical model.

\subsubsection{Results for the high-energy acceleration chain}

The same study is then extended to the remaining RCSs. The 8-arc lattices of RCS2 and RCS3 also exhibit significant sensitivity to the energy--magnetic-field mismatch, with horizontal emittance growths of \qty{30}{\percent} and \qty{22}{\percent}, respectively. For this reason, and following the conclusions of the longitudinal dynamics studies, including intensity effects \cite{thiele2026modelling}, the latest lattice design adopts 32 RF stations in each RCS. The results of the energy--magnetic-field mismatch study for this RF distribution are summarised in Table~\ref{tab:EB_rcs_chain}.
For RCS1, the 32-arc lattice exhibits a reduced trajectory excursion of \qty{4}{\milli\meter}, compared with \qty{13}{\milli\meter} for the 8-arc lattice. The beta-beating is also reduced from \qty{20}{\percent} to approximately \qty{3}{\percent}, highlighting the role of a more distributed RF system in reducing the perturbation. The horizontal emittance growth is reduced to \qty{7}{\percent}. Although this represents a significant improvement, it still accounts for most of the transverse emittance growth budget allocated to the full acceleration chain, from the exit of the cooling channel to the collider, which is \qty{10}{\percent}.
The downstream RCSs are less critical, as they operate at higher energies and with lower magnetic-field ramp rates than RCS1. The horizontal emittance growth is of the order of a few percent in each of the remaining RCSs, while the trajectory excursion remains of the order of \qty{1}{\milli\meter}. Despite this improvement, the resulting perturbations still need to be corrected to ensure acceptable beam transport through the acceleration chain.

\begin{table}
\caption{\label{tab:EB_rcs_chain} Impact of the energy–magnetic-field mismatch on the beam trajectory, optical functions, and emittance growth across the RCS chain for the 32-arc lattices.}
\begin{ruledtabular}
\begin{tabular}{ccccc}
Parameter & RCS1 & RCS2 & RCS3 & RCS4 \\
\hline
Max. $\Delta x$ [mm] 
& 4.1 & 0.5 & 0.4 & 1 \\
Max. $\Delta\beta_x/\beta_x$ [\%]
& 3.3 & 0.4 & 0.3 & 1 \\
Max. $\Delta\beta_y/\beta_y$ [\%]
& 1.4 & 0.5 & 0.1 & 3 \\
\hline
Final $\Delta\varepsilon_x/\varepsilon_x$ [\%]
& 6.8 & 0.4 & 1 & 1.5 \\
Final $\Delta\varepsilon_y/\varepsilon_y$ [\%]
& -0.2 & -0.3 & -0.1 & 0.5 \\
Final $\Delta\varepsilon_s/\varepsilon_s$ [\%]
& 3.4 & 2.8 & -0.5 & 1.5 \\
\end{tabular}
\end{ruledtabular}
\end{table}

\subsubsection{Perspectives of correction}

The correction of this error is particularly challenging in the presence of counter-rotating $\mu^+$ and $\mu^-$ beams sharing the same beam pipe. Magnetic dipole correctors or magnet sectoring would require correction fields of opposite signs for the two beams, such that correcting the field error for one beam would worsen it for the other. Electrostatic correctors can provide corrections to both beams, but would require unrealistically high voltages of the order of \unit{\giga\volt}. Alternative approaches, such as exploiting a field index in the dipoles, could be promising but require further investigation.

\section{Impact of a ramp jitter}

Given the unprecedented ramp rate, ramp jitter is a concern, as the power supply must deliver a fast current ramp with high pulse-to-pulse repeatability. The main source of error is the switching-time jitter of the electronic switches in the power cells. The power supply repeatability requirement specifies that the current deviation from the average ramp, \(\Delta I\), remains below \qty{100}{ppm} for approximately \qty{95}{\percent} of pulses, corresponding to a $2\sigma$ tolerance \cite{powering_FBoattini_2025}. The following evaluates the impact of ramp jitter on beam dynamics for different error amplitudes, considering worst-case scenarios. Rather than scaling the error with the instantaneous current, a constant deviation, $\Delta I$, set by the jitter amplitude and expressed as a fraction of the extraction current is assumed, providing a conservative upper bound on the expected error.
Let $I_\text{inj}$, $I_\text{ext}$, and $I_0(t)$ denote the nominal injection and extraction currents and nominal ramp, respectively, and $t$ and $t_{acc}$ the instantaneous time and total acceleration time, respectively. To assess the worst-case impact of jitter, four current-ramp scenarios are considered:
\begin{equation}
    \begin{aligned}
    I_0(t) &= I_\text{inj} + \left(I_\text{ext} - I_\text{inj}\right) \frac{t}{t_\text{acc}} \\
    I_{--}(t) &= I_0(t) - \Delta I\\
    I_{-+}(t) &= I_0(t) + \Delta I \left(2\frac{t}{t_\text{acc}}-1\right) \\
    I_{++}(t) &= I_0(t) + \Delta I\\
    I_{+-}(t) &= I_0(t) - \Delta I \left(2\frac{t}{t_\text{acc}}-1\right) 
\end{aligned}
\end{equation}

The first ramp $I_{--}$ remains uniformly below the nominal ramp by $\Delta I$, while $I_{++}$ remains uniformly above it by the same amount. Ramp $I_{-+}$ starts below the nominal ramp and finishes above it. Conversely, $I_{+-}$ starts above and ends below the ramp.

\subsection{Orbit and optics perturbations}

\begin{figure*}
    \centering
    \begin{subfigure}{0.45\textwidth}
        \centering
        \includegraphics[width=\linewidth]{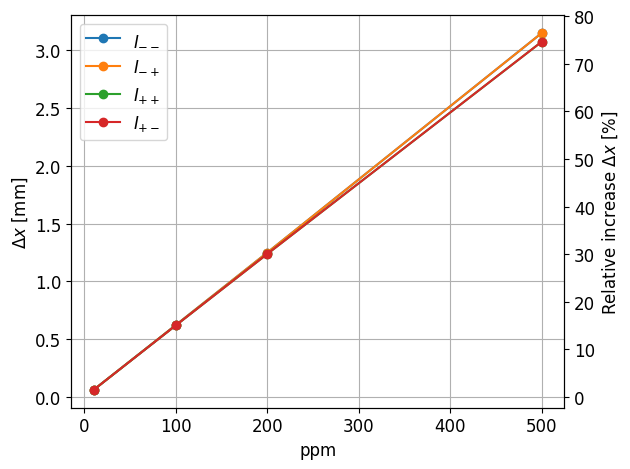}
        \caption{$\Delta x$.}
        \label{subfig:jitter_D_x}
    \end{subfigure}
    \hfill
    \begin{subfigure}{0.45\textwidth}
        \centering
        \includegraphics[width=\linewidth]{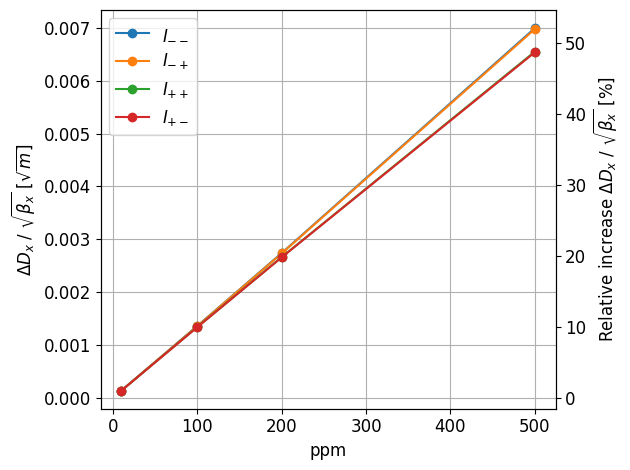}
        \caption{$\Delta D_x / \sqrt{\beta_x}$.}
        \label{subfig:jitter_D_dx}
    \end{subfigure}
    \begin{subfigure}{0.45\textwidth}
        \centering
        \includegraphics[width=\linewidth]{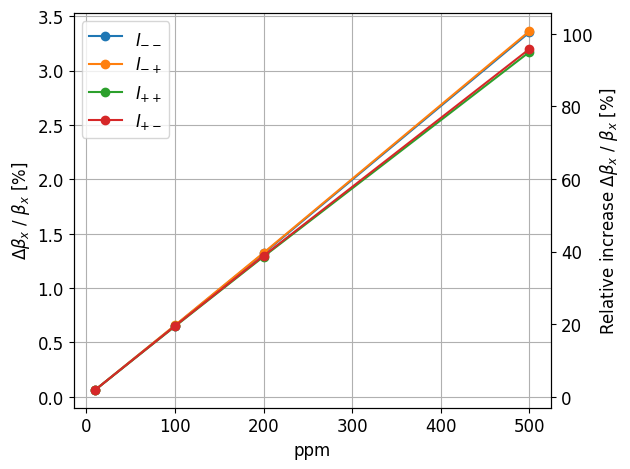}
        \caption{$\Delta \beta_x / \beta_x$.}
        \label{subfig:jitter_D_betax}
    \end{subfigure}
    \hfill
    \begin{subfigure}{0.45\textwidth}
        \centering
        \includegraphics[width=\linewidth]{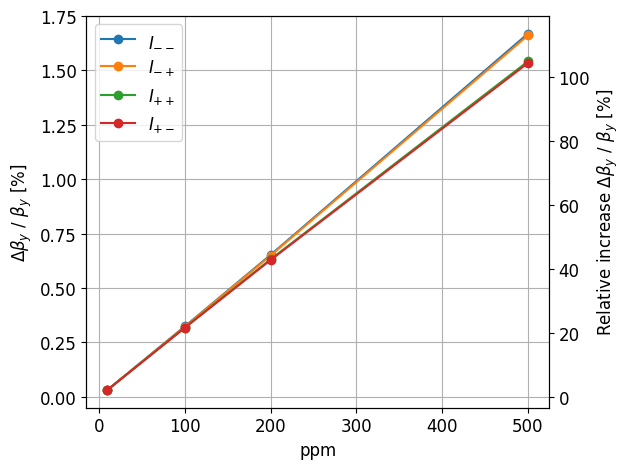}
        \caption{$\Delta \beta_y / \beta_y$.}
        \label{subfig:jitter_D_betay}
    \end{subfigure}
    \caption{Relative increase in the maximum trajectory offset and optics perturbations compared with the nominal ramp for the RCS1 32-arc lattice, considering different jitter amplitudes and worst-case scenarios.}
    \label{fig:jitter_Doptics}
\end{figure*}

The jitter is treated as a perturbation of the nominal fast ramp. Throughout this section, the subscripts \textit{reference}, \textit{nominal}, and \textit{jitter} refer to the matched optics, nominal fast ramp, and dephased fast ramp, respectively. We investigate the impact of jitter on the orbit and optical functions in the 32-arc RCS1 lattice. The baseline maximum deviations from the matched optics for the nominal fast ramp are reported in Table~\ref{tab:EB_rcs_chain}.
The impact of each jitter scenario is quantified by the increase in orbit distortion and optics perturbations relative to the nominal ramp, calculated as follows:
\begin{equation}
    \begin{aligned}
    \frac{\Delta \beta}{\beta} &= \frac{\beta_\text{jitter} - \beta_\text{nominal}}{\beta_\text{reference}} \\
    \frac{\Delta D_x}{\sqrt{\beta_x}} &= \frac{D_\text{x, jitter} - D_\text{x, nominal}}{\sqrt{\beta_\text{x, reference}}} \\
    \Delta x &= x_\text{jitter} - x_\text{nominal}
\end{aligned}
\end{equation}
These values are normalized to the nominal ramp perturbations reported in Table~\ref{tab:EB_rcs_chain}, to quantify the relative increase due to jitter. The results, shown in Fig.~\ref{fig:jitter_Doptics}, exhibit a linear dependence on the jitter amplitude. This is expected for a linear lattice composed only of dipoles and quadrupoles. At higher jitter amplitudes of \qty{500}{ppm}, the ramp scenarios $I_{--}$ and $I_{-+}$ create the largest perturbations for all quantities compared to the nominal ramp case. These scenarios exhibit larger errors at the arc entrance than at its exit, suggesting that the initial mismatch at injection may contribute to the magnitude of the final perturbations.
For a tolerance of \qty{100}{ppm}, the maximum orbit distortion increases by slightly more than \qty{0.5}{\milli\meter}, corresponding to approximately \qty{15}{\percent} of the nominal ramp offset. This increase reaches about \qty{75}{\percent} for a \qty{500}{ppm} jitter, or \qty{3}{\milli \meter}. In absolute values, the horizontal $\beta$-beating is affected more strongly than the vertical $\beta$-beating, due to the horizontal off-axis passage of the beam through the quadrupoles. For a \qty{100}{ppm} jitter, the $\beta$-beating in both transverse planes increases by approximately \qty{20}{\percent}, reaching an increase of \qty{100}{\percent} for a \qty{500}{ppm} jitter. The horizontal dispersion perturbation increases by \qty{10}{\percent} for a \qty{100}{ppm} jitter and by \qty{50}{\percent} for a \qty{500}{ppm} jitter.

\subsection{Tracking studies and emittance growth}

\begin{figure*}
    \centering
    \begin{subfigure}{0.48\textwidth}
        \centering
        \includegraphics[width=\linewidth]{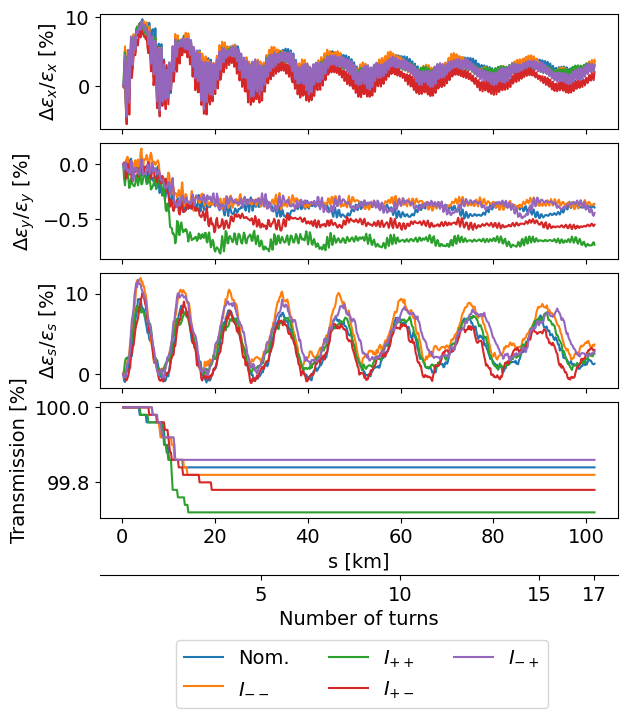}
        \caption{\qty{100}{ppm} jitter.}
        \label{subfig:tracking_100ppm}
    \end{subfigure}
    \hfill
    \begin{subfigure}{0.48\textwidth}
        \centering
        \includegraphics[width=\linewidth]{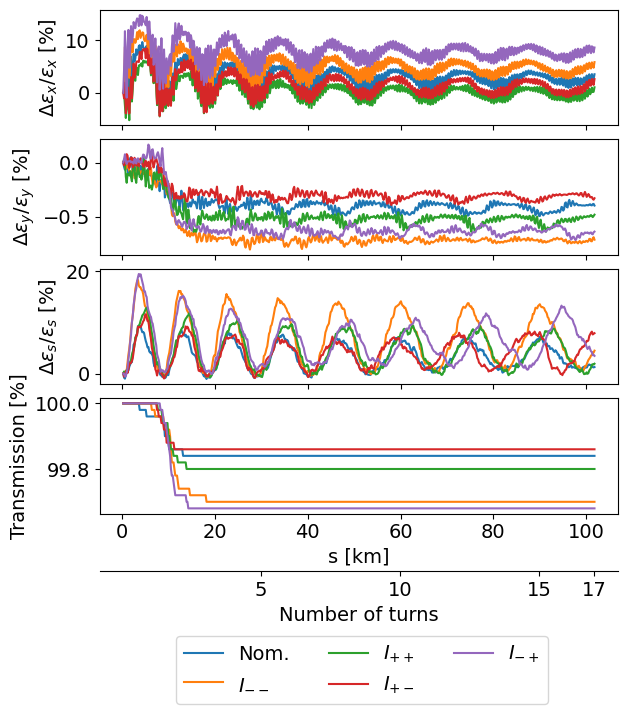}
        \caption{\qty{500}{ppm} jitter.}
        \label{subfig:tracking_500ppm}
    \end{subfigure}
    \caption{Tracking results for the RCS1 32-arc lattice, considering different ramp scenarios and jitter amplitudes of \qty{100}{ppm} and \qty{500}{ppm}.}
    \label{fig:tracking_jitter}
\end{figure*}

A final study investigates the impact of ramp jitter on beam quality and emittance preservation. For each ramp scenario, the jitter amplitude was scanned, and the resulting emittance growth was compared with that obtained for the nominal ramp. The results are presented in Fig.~\ref{fig:tracking_jitter}. For jitter amplitudes up to \qty{100}{ppm}, corresponding to the expected tolerance, the additional emittance growth remains negligible compared with the nominal case. The slight decrease in vertical emittance results from the loss of particles at the beginning of the acceleration cycle, which effectively reduces the emittance of the surviving beam.

At a higher jitter amplitude of \qty{500}{ppm}, differences emerge between the ramp scenarios. Interestingly, the $I_{+-}$ and $I_{++}$ ramps lead to reduced horizontal emittance growth compared with the nominal ramp, whereas the $I_{--}$ and $I_{-+}$ ramps increase it. This behaviour is consistent with the previously observed optics perturbations: the $I_{+-}$ and $I_{++}$ scenarios reduce the field error at the beginning of the arc, suggesting that the initial mismatch at injection may contribute significantly to the magnitude of the final perturbations. Similarly, the $I_{--}$ and $I_{-+}$ ramps lead to stronger perturbations of the longitudinal emittance: the amplitude of the longitudinal emittance oscillations, driven by the initial mismatch at injection, increases compared with the other ramp scenarios and the nominal case. 

\section{Conclusion and perspectives}

RCS1 is the most critical ring of the acceleration chain due to its extremely fast ramp. The combination of strong synchro-betatron coupling, large longitudinal emittance, and relatively low beam energy makes an \num{8}-section RF layout insufficient to preserve beam quality. Increasing the RF distribution to \num{32} sections significantly improves beam transport, although second-order optical matching remains necessary to suppress residual emittance growth. \par
The mismatch between the beam energy and the magnetic fields of the pulsed elements induces orbit distortions, significant $\beta$-beating, and substantial emittance growth. A more distributed RF system mitigates these effects, with \num{32} RF sections limiting horizontal emittance growth to approximately \qty{7}{\percent} in RCS1 and less than \qty{1}{\percent} in RCS2–RCS4. The growth in RCS1 consumes a substantial fraction of the overall \qty{10}{\percent} emittance budget for the acceleration complex \cite{IMCC_interimreport2025}. An innovative correction scheme is therefore required to mitigate the effects of the energy–magnetic-field mismatch and limit its impact on beam dynamics. \par
Ramp jitter was also investigated, with beam perturbations scaling linearly with jitter amplitude. At the nominal power-supply specification of \qty{100}{ppm}, the impact on emittance growth is limited compared with that of the nominal fast ramp. \par
Further studies should assess the robustness of the lattice against realistic magnetic-field errors, alignment imperfections, and power-supply fluctuations. The development of correction schemes and the definition of associated tolerances are essential next steps. Finally, start-to-end simulations of the complete acceleration chain are required to verify that the transverse emittance budget can be maintained when all relevant machine imperfections are included. \par

\FloatBarrier
\appendix
\begin{widetext}
\section{Analytical expressions for the trajectory offset and matrix perturbation}\label{sec:appendix}

\subsection{Pulsed dipole}
The integrated orbit offset $\Delta X_{\text{dip}}$ generated by a pulsed dipole is given by:
\begin{align}
\Delta X_{dip}(L) = k'_0
\begin{pmatrix}
\dfrac{-k_0^* (L + s_0 - s^*) + k_0^* (s_0 - s^*) \cos(k_0^* L) + \sin(k_0^* L)}{k_0^{*3}} \\[1em]
\dfrac{ -1 + \cos(k_0^* L) - k_0^* (s_0 - s^*) \sin(k_0^* L)}{k_0^{*2}} \\[1em]
0 \\[0.5em]
0 \\[0.5em]
\dfrac{-2 + k_0^{*2} L (L + 2 s_0 - 2 s^*) + 2 \cos(k_0^* L) + 2 k_0^* (-s_0 + s^*) \sin(k_0^* L)}{2 k_0^{*3} \beta_0} \\[1em]
0
\end{pmatrix}
\end{align}
The non-zero elements of the integrated matrix perturbation $\Delta M_{\text{dip}}$ are:
\begin{align}
\Delta M_{11}
&= - \frac{k'_0}{2 k_0^{*2}}
\Bigl[
k_0^*L\cos(k_0^*L)
+ \sin(k_0^*L)
\bigl[
1 - 2\cos(k_0^*L)
+ 2k_0^*(s_0-s^*)\sin(k_0^*L)
\bigr]
\Bigr]
\\
\Delta M_{12}
&= - \frac{k'_0}{2 k_0^{*3}}
\Bigl[
-2\cos(k_0^*L)
+ 2\cos^2(k_0^*L)
+ k_0^*
\Bigl(
L
+ 2(s_0-s^*)
\bigl[
1-\cos(k_0^*L)\sin(k_0^*L)
\bigr]
\Bigr)
\Bigr]
\\
\Delta M_{16}
&= \frac{k'_0}{2 k_0^{*3}\beta_0}
\Bigl[
-k_0^*(2L+s_0-s^*)
+ k_0^*(L+2s_0-2s^*)\cos(k_0^*L)
\notag\\
&\qquad\qquad
-k_0^*(s_0-s^*)\cos(2k_0^*L)
+ 3\sin(k_0^*L)
-\sin(2k_0^*L)
\Bigr]
\\
\Delta M_{21}
&= \frac{k'_0}{2}
(L+2s_0-2s^*)
\sin(k_0^*L)
\\
\Delta M_{22}
&= \frac{k'_0}{2k_0^{*2}}
\Bigl[
k_0^*L\cos(k_0^*L)
-\sin(k_0^*L)
\Bigr]
\\
\Delta M_{26}
&= \frac{k'_0}{2k_0^{*2}\beta_0}
\Bigl[
-2
+2\cos(k_0^*L)
+k_0^*L\sin(k_0^*L)
\Bigr]
\\
\Delta M_{34}
&= \frac{k'_0}{2k_0^{*3}}
\Bigl[
2
-k_0^{*2}L(L+2s_0-2s^*)
-2\cos(k_0^*L)
+2k_0^*(s_0-s^*)\sin(k_0^*L)
\Bigr]
\\
\Delta M_{51}
&= \frac{k'_0}{2k_0^{*2}\beta_0}
\Bigl[
-2
+2\cos(k_0^*L)
+k_0^*L\sin(k_0^*L)
\Bigr]
\\
\Delta M_{52}
&= \frac{k'_0}{2k_0^{*3}\beta_0}
\Bigl[
2k_0^*(s_0-s^*)
-k_0^*(L+2s_0-2s^*)\cos(k_0^*L)
+\sin(k_0^*L)
\Bigr]
\\
\Delta M_{56}
&= - \frac{k'_0}{2k_0^{*3}(\gamma_0^2-1)}
\Bigl[
-2
+k_0^{*2}L(L+2s_0-2s^*)
+2\cos(k_0^*L)
\notag\\
&\qquad\qquad
-2k_0^*(s_0-s^*)\sin(k_0^*L)
+k_0^*\gamma_0^2(L+2s_0-2s^*)
\bigl[
-k_0^*L+\sin(k_0^*L)
\bigr]
\Bigr]
\end{align}

\subsection{Pulsed quadrupole}
A pulsed quadrupole does not generate an orbit offset. Therefore,
\begin{align}
    \Delta X_{qd}=0
\end{align}
For a focusing quadrupole ($k_1 > 0$), the non-zero elements of the integrated matrix perturbation $\Delta M_{\text{qd}}$ are:
\begin{align}
\Delta M_{11}
&= - \frac{k'_1}{4 k_1^{3/2}}
\Bigl[
\sqrt{k_1}L\cos(\sqrt{k_1}L)
+
\bigl[
-1 + k_1 L (L + 2 s_0 - 2 s^*)
\bigr]
\sin(\sqrt{k_1}L)
\Bigr]
\\
\Delta M_{12}
&= \frac{k'_1}{4 k_1^{3/2}}
\Bigl[
(L + 2 s_0 - 2 s^*)
\bigl[
\sqrt{k_1}L\cos(\sqrt{k_1}L)
- \sin(\sqrt{k_1}L)
\bigr]
\Bigr]
\\
\Delta M_{21}
&= - \frac{k'_1}{4 \sqrt{k_1}}
\Bigl[
(L + 2 s_0 - 2 s^*)
\bigl[
\sqrt{k_1}L\cos(\sqrt{k_1}L)
+ \sin(\sqrt{k_1}L)
\bigr]
\Bigr]
\\
\Delta M_{22}
&= - \frac{k'_1}{4 k_1^{3/2}}
\Bigl[
-\sqrt{k_1}L\cos(\sqrt{k_1}L)
+
\bigl[
1 + k_1 L (L + 2 s_0 - 2 s^*)
\bigr]
\sin(\sqrt{k_1}L)
\Bigr]
\\
\Delta M_{33}
&=
\frac{k'_1}{4 k_1^{3/2}}
\Bigl[
-\sqrt{k_1}L\cosh(\sqrt{k_1}L)
+
\bigl(1+k_1L(L+2s_0-2s^*)\bigr)
\sinh(\sqrt{k_1}L)
\Bigr]
\\
\Delta M_{34}
&=
\frac{k'_1}{4 k_1^{3/2}}
(L+2s_0-2s^*)
\Bigl[
\sqrt{k_1}L\cosh(\sqrt{k_1}L)
-
\sinh(\sqrt{k_1}L)
\Bigr]
\\
\Delta M_{43}
&=
\frac{k'_1}{4\sqrt{k_1}}
(L+2s_0-2s^*)
\Bigl[
\sqrt{k_1}L\cosh(\sqrt{k_1}L)
+
\sinh(\sqrt{k_1}L)
\Bigr]
\\
\Delta M_{44}
&=
\frac{k'_1}{4k_1^{3/2}}
\Bigl[
\sqrt{k_1}L\cosh(\sqrt{k_1}L)
+
\bigl(-1+k_1L(L+2s_0-2s^*)\bigr)
\sinh(\sqrt{k_1}L)
\Bigr]
\end{align}
For a defocusing quadrupole, the corresponding expressions are obtained by replacing $k_1$ and $k'_1$ with $- k_1$ and $- k'_1$, exchanging the horizontal and vertical blocks, and reversing the appropriate signs.

\end{widetext}

\begin{acknowledgments}
Endorsed by the IMCC. Funded by the European Union (EU). Views and opinions expressed are however those of the author(s) only and do not necessarily reflect those of the EU or European Research Executive Agency (REA). Neither the EU nor the REA can be held responsible for them.
\end{acknowledgments}

\bibliography{bib}

\end{document}